# Magneto-Ionic Vortices: Voltage-Reconfigurable Swirling-Spin Analog-Memory Nanomagnets

Irena Spasojevic[1,*], Zheng Ma[1], Aleix Barrera[2], Federica Celegato[3], Ana Palau[2], Paola Tiberto[3], Kristen S. Buchanan[4], Jordi Sort[1,5,*]

[1]Departament de Física, Universitat Autònoma de Barcelona, 08193 Bellaterra, Spain
[2]Institut de Ciència de Materials de Barcelona (ICMAB-CSIC), Campus UAB, Bellaterra 08193, Barcelona, Spain
[3]Advanced materials and Life science Divisions, Istituto Nazionale di Ricerca Metrologica (INRIM), Strada delle Cacce 91, 10135 Turin, Italy
[4]Department of Physics, Colorado State University, Fort Collins, Colorado 80523, USA
[5]ICREA, Pg. Lluís Companys 23, 08010 Barcelona, Spain

**Abstract**

Rapid progress in information technologies has spurred the need for innovative memory concepts, for which advanced data-processing methods and tailor-made materials are required. Here we introduce a previously unexplored nanoscale magnetic object: an analog magnetic vortex controlled by electric-field-induced ion motion, termed magneto-ionic vortex or "vortion". This state arises from paramagnetic FeCoN through voltage gating and gradual N3- ion extraction within patterned nanodots. Unlike traditional vortex states, vortions offer comprehensive analog adjustment of key properties such as magnetization amplitude, nucleation/annihilation fields, or coercivity using voltage as an energy-efficient tuning knob. This manipulation occurs post-synthesis, obviating the need for energy-demanding methods like laser pulses or spin-torque currents. By leveraging an overlooked aspect of N3- magneto-ionics -planar ion migration within nanodots- precise control of the magnetic layer's thickness is achieved, which enables reversible transitions among paramagnetic, single-domain, and vortion states, offering future prospects for analog computing, multi-state data storage, or brain-inspired devices.

[*]Email: Irena.Spasojevic@uab.cat, Jordi.Sort@uab.cat

With the advent of Big Data, energy resources spent on information technologies are growing exponentially[1]. One of the reasons for this is that most memory systems utilize electric currents to write data, which dissipates power by Joule heating[2]. Control of magnetic memories with electric fields instead of electric currents has emerged as a leading strategy to minimize this problem[3-6]. Among the various mechanisms to tune magnetism with voltage (electrostatic charging, strain-mediated multiferroic coupling or electrochemical reactions[3,7]), magneto-ionics, which involves the manipulation of magnetic properties through voltage-driven insertion/removal of ions, provides unprecedented performance for non-volatile control of coercivity, anisotropy, exchange bias or magnetization, ultimately causing conversion between magnetic and non-magnetic states[8-14]. So far, most investigated magneto-ionic systems are continuous thin films utilizing different ion species ($O^{2-}$, $N^{3-}$, $H^+$, $Li^+$, etc.). Reports on magneto-ionic effects in nanoscale lithographed dots are scarce[15,16]. From the fundamental viewpoint, *in-situ* measurements of voltage-driven ion motion in nanometer-sized structures might unveil changes in the magnetization reversal mechanisms, which may allow precise, analog control of magnetic bit properties that can be used to create novel neuromorphic functionalities.

At the nanoscale, unique magnetization switching mechanisms and topological spin configurations emerge, such as skyrmions or vortices. Magnetic vortices hold promise in applications such as multi-state data storage[17], racetrack memories[18], biomedicine[19], spin-logics[20], or spin-torque nano-oscillators[21], where the latter is relevant to microwave generation/detection[22] and neuromorphic computing[23]. Magnetic vortices compete with other spin configurations such as single-domain (SD), C- or S-states, or double-vortices[24,25]. The ground state of the system is determined by the interplay between exchange and magnetostatic energy and is traditionally controlled by tuning the aspect ratio of as-grown nanostructures according to existing phase diagrams. This implies that altering the magnetization reversal mechanism and the magnetic properties of nanodots after sample growth is not straightforward. Post-synthesis manipulation of magnetic vortices has been accomplished, to some extent, by spin-torque effects[26], laser pulse excitation[27], or through the coupling with adjacent antiferromagnets[28], all requiring significant energy resources. Consequently, alternative approaches to controlling magnetism in nanoscale objects, by harnessing electric fields rather than relying on electrical currents or heating processes, are needed.

Voltages applied across piezoelectric substrates with magnetostrictive nanodots grown on top can trigger the creation and annihilation of magnetic vortices[29], a change of the vortex chirality/polarity[30], or a magnetization reorientation in SD nanodots[31] *via* the Villari effect. However, strain-mediated coupling cannot induce an analog modulation of the magnetization amplitude, crucial for mimicking weights in artificial synapses,[32] nor can it be used to precisely control the nucleation/annihilation fields of vortex states. Furthermore, strain-coupled devices suffer from long-term fatigue and eventual cracking that limit their useful lifetime. Moreover, the magnetic state of each nanodot is often unpredictably influenced by the local strain stemming from the complex ferroelectric domain structure of the substrate. Considering all the shortcomings of existing strategies, the analog modification of magnetic properties *via* voltage-induced migration of ions in nanometer-sized dots arises as an unparalleled alternative.

Here, we demonstrate the operation principle of an analog magnetic *vort*ex controlled by electric-field-driven *ion* motion ("*vortion*" or *magneto-ionic vortex*). Such state is generated from paramagnetic FeCoN nanodots upon voltage-driven gradual extraction of $N^{3-}$ ions. The characteristic properties of vortions (nucleation and annihilation fields, coercivity, or remanence) can be fully controlled with voltage. While these objects show some resemblance to conventional magnetic vortices, a unique feature of vortions is that the amplitude of their magnetization, akin to synaptic weight, can be dynamically reconfigured in an analog manner by adjusting the voltage actuation time. Additionally, voltage can induce a reversible transition among three distinct states *via* migration of $N^{3-}$ ions – paramagnetic, single-domain and magneto-ionic vortex– which is not possible using conventional methods. Such control exploits a rather unique, so far overlooked, aspect of $N^{3-}$ magneto-ionics: contrary to $O^{2-}$ magneto-ionics (where voltage generates ferromagnetic clusters within a paramagnetic matrix), diffusion of $N^{3-}$ ions results in the formation of a planar migration front, dividing the nanodot into sub-layers with dissimilar composition and magnetic features.

**Unraveling the Magneto-Ionic Vortex: formation and control**

FeCoN nanodot arrays were prepared by electron beam lithography, followed by magnetron sputtering and lift-off, as depicted in **Fig. 1a** (see Methods). Atomic force microscopy (AFM) topography and transmission electron microscopy (TEM) images show that patterned nanodots have diameter of 280 nm and thickness around 35 nm (**Fig. 1c** and **Supplementary Fig. 1**). Energy dispersive X-ray analysis (EDX) reveals a Fe:Co atomic ratio of 65:35 (**Fig. 1d**). This composition was chosen since it shows the largest nitrogen magneto-ionic effects reported so far[33]. To trigger ionic migration, we employed an electrolyte-gated setup in a capacitor configuration within a custom-made electrochemical cell (**Fig. 1b**) that allowed us to track real-time alterations of magnetic properties using magneto-optical Kerr effect (MOKE). Propylene carbonate (PC) with $Na^+$- and $OH^-$- solvated species was employed as an anhydrous polar liquid electrolyte. Its role is twofold: to act as a reservoir for $N^{3-}$ ions, and to facilitate the creation of an electric double layer (EDL), which comprises a sub-nm-thick layer of solvated ions drawn to the interface between the gated layer and the electrolyte. This generates a strong electric field capable of efficiently transporting $N^{3-}$ ions into and out of FeCoN, depending on the gate voltage ($V_G$) polarity[6]. The as-grown FeCoN nanodots are paramagnetic with virtually no Kerr amplitude (**Fig. 1e**, purple line, and **Supplementary Fig. 2a**). Upon application of $V_G = -25$ V, $N^{3-}$ ions begin to migrate towards the PC, generating ferromagnetic phases within each FeCoN nanodot. After 6 min, an appreciable, narrow hysteresis loop emerges, characterized by a Kerr amplitude of 0.5 mdeg (**Supplementary Fig. 2b**). Interestingly, with increasing actuation time, not only does the Kerr amplitude increase, but the shape of hysteresis loops undergoes a transformation, from square-like (short actuation times) to constricted loops (longer actuation times) (**Supplementary Fig. 2c,d**), typical of vortex formation. When a voltage of opposite polarity is applied ($V_G = +25$ V), $N^{3-}$ ions are reintroduced from the electrolyte into the FeCo(N) layer. This results in a decrease of the Kerr amplitude (**Fig. 1f**), accompanied by a change of hysteresis loops' shape (from vortex-like to square-like) and, finally, recovery of the paramagnetic FeCoN state (**Supplementary Fig. 2e-h**).

The dependance of the Kerr signal amplitude at saturation ($A_{Kerr}$), squareness ratio (*i.e.*, remanence-to-saturation ratio, $M_R/M_S$), saturation field (annihilation for the vortex state) ($H_S$ or $H_A$), vortex nucleation field ($H_N$) and coercivity ($H_C$) on time is plotted in **Fig 2**. The blue curves illustrate the time evolution of specific parameters under $-25$ V. In contrast, the red curves depict the change in the same parameters over time under + 25 V, providing information about the reversibility of magneto-ionic actuation. As shown in **Fig. 2a,** $A_{Kerr}$ progressively increases in the presence of a negative voltage, and diminishes when the voltage polarity is reversed, as expected. The $M_R/M_S$ ratio (**Fig. 2b**) exhibits a non-monotonic behavior while applying – 25 V. It first increases with time while the loops present no constriction (SD state) but, after 15 min, $M_R/M_S$ starts to decrease for actuation times up to 50 min, after which a clear loop constriction is observed. Notably, this behavior follows the same trend upon the application of + 25 V. The abrupt change in $M_R/M_S$ at 50 min marks the onset in the emergence of a certain nucleation field and coincides with a sudden alteration in $H_A$ characteristics, as illustrated in **Fig. 2c**. Once again, the behaviors of $H_S$ (or $H_A$) and $H_N$ follow a reversible and opposite trend when a positive voltage is applied. The concurrent alterations in the mentioned parameters align with the emergence of clear constriction in the hysteresis loops and therefore indicate a change in the magnetization reversal mechanism—from coherent rotation (orange region, with square-like loops) to vortex formation (blue region). The transition from coherent rotation to vortex states and *vice versa* proceeds gradually, as denoted by grey shadowed regions in **Fig. 2a-d**. Within these transient regimes, the hysteresis loops' shape falls somewhere between the two states (**Supplementary Fig. 2c,f**)). The black dotted lines in **Fig. 2a-d** delimitate the regions where vortions are stable for positive and negative voltage actuation, respectively.

Remarkably, upon application of $V_G = -25$ V, once in the vortion regime, both $H_A$ and $H_N$ clearly evolve with time, where increasing voltage actuation time dynamically impacts the stability of the vortex state. Specifically, when coming from positive saturation (descending branch of the hysteresis loops), extending the voltage actuation time results in the vortex state nucleating earlier and annihilating later. Thus, longer application of voltage leads to increased stability of the vortex state.

Our observations are consistent with an increase in thickness of the formed ferromagnetic phase as depicted in **Fig. 2e**. Commencing with paramagnetic FeCoN, brief negative voltage actuation induces the formation of a thin ferromagnetic layer within FeCoN nanodots, giving rise to a SD state and magnetization reversal through coherent rotation. This layer could be FeCo alloy or $(Fe,Co)_4N$, both of which are ferromagnetic. With prolonged negative voltage application, the thickness of the ferromagnetic layer gradually increases, destabilizing the SD state and causing the nucleation of the vortex state[25]. The progressive increase in the thickness of the ferromagnetic layer further stabilizes the vortex configuration (thereby increasing $H_N$ and $H_A$), in agreement with the literature[34,35]. Importantly, in our case, magneto-ionic actuation allows precise tuning of the magnetic phase thickness within the same sample. This allows accurate post-synthesis control over the magnetization reversal mechanism and enables analog modulation of magnetization along with the other characteristic parameters of the magneto-ionic vortex state.

We delved deeper into the impact of actuation time on the magnetization reversal mechanism by MFM imaging. **Fig. 3** displays two hysteresis loops obtained by MOKE after voltage treatments for varying durations, together with overlaid MFM phase images collected at different values of in-plane magnetic field. Results in panel a) were obtained for short actuation times, giving rise to magnetization reversal *via* coherent rotation (*i.e.*, rotating dipolar contrast as a function of magnetic field). Conversely, for long-term actuation (panel b), one can observe magnetization reversal *via* vortex formation (MFM phase featureless contrast around coercivity), in agreement with the constriction observed in the hysteresis loop. As expected, no contrast was observed in the as-grown state (**Supplementary Fig. 3a**).

**Fig. 2d** reveals that $H_C$, during − 25 V gating, exhibits a minor peak precisely at the moment when the vortex state starts to form. An analogous behavior is evident in the opposite case scenario, albeit with significantly higher $H_C$ values. Enhancement of $H_C$ at the transition between SD and vortex states has been previously reported.[36] Superimposed to this behavior is a steady increase of $H_C$, observed while going from SD to vortex state and *vice versa*, which is persistent over time. The increase of $H_C$ under negative voltage is consistent with the formation of FeCo or (Fe,Co)$_4$N magnetic phases. As previously observed in FeCoN films, a reduction in nitrogen content (*i.e.*, from FeCoN$_x$ to FeCo) results in progressively higher magnetization and coercivity[37,38]. The final coercivity values during positive gating ($\approx$ 170 Oe) are substantially higher compared to initial values ($\approx$ 50 Oe) when negative voltage is applied, indicating the presence of magnetic phases with dissimilar nitrogen content at the beginning and at the end of gating with negative and positive voltages, respectively.

**Microstructural insights into the mechanisms of ion migration**

To elucidate the ion migration mechanism in FeCoN nanodots, we conducted high-resolution TEM (HR-TEM), high-angle annular dark-field scanning transmission electron microscopy (HAADF-STEM) and electron energy loss spectroscopy (EELS) compositional mapping on cross-sectional lamellae extracted from as-grown and − 25 V treated samples. EELS compositional maps for each sample state were collected in two distinct regions, as indicated by rectangles in HAADF-STEM images of the nanodot's cross-sections (**Fig. 4a,d**), covering the entire nanodot and its interior. Additionally, for each state, we measured EELS compositional line profile across the nanodot (**Fig. 4g,h**). A uniform distribution of Fe, Co, and N ions within the dot is observed in the as-grown sample (**Fig. 4b,c,g**). Following the application of negative voltage, ions within the sample exhibit a distinct, altered distribution. Specifically, positively charged Fe ions migrate towards the bottom interface with Pt, where they accumulate (**Fig. 4f,h**, blue line). Upon treatment, the concentration of Co ions also slightly increases towards the bottom interface of the nanodot. Similar motion of $Ni^{2+}$ ions upon voltage application has been recently reported in gated NiO films[39]. A more intriguing aspect lies in the distinct behavior of negatively charged $N^{3-}$ ions, exhibiting a counter-directional migration compared to transition metal ions, that following partial release into the electrolyte, also accumulate at the top interface of the FeCoN films. This generates a distinctive planar migration front which subdivides the FeCoN dot into two sub-layers with dissimilar nitrogen content: N-depleted or N-free (at the bottom) and N-rich (top) of the nanodot (**Fig. 4f,h**). This nitrogen accumulation phenomenon has been observed in systems characterized by relatively fast nitrogen magneto-ionic effects. When nitrogen diffuses faster than it dissolves, the EDL becomes saturated with $N^{3-}$ ions, locally surpassing the solubility limit of nitrogen in PC. In this scenario, $N^{3-}$ ions reach a point where they cannot be released further into the liquid, resulting in their accumulation at the upper part of the FeCoN nanodot[33,40].

Our observations underscore the critical significance of the formed nitrogen front in modulating the thickness of the induced ferromagnetic counterpart by adjusting the voltage actuation time. This phenomenon, absent in oxygen magneto-ionic systems such as CoO$_x$ and FeO$_x$ [12,16], plays a pivotal role in the transition from a SD to a vortion state. In most other systems, ion migration leads to the emergence of ferromagnetic clusters within a paramagnetic matrix, as opposed to the formation of distinct ferromagnetic thin bottom sub-layer alongside paramagnetic top sub-layer.

HR-TEM images of nanodot cross-sections and fast Fourier transforms (FFTs) of the areas enclosed within orange squares in **Fig. 4i,j**, were utilized to elucidate the nature of the phases before and after gating at – 25 V, respectively. As-grown FeCoN nanodot has a nanocrystalline structure and exhibits two sets of discrete spots, corresponding to interplanar distances of 2.26 Å (yellow circles) and 2.60 Å (cyan circles), consistent with (200) and (111) planes of face-centered cubic (fcc) $F\bar{4}3m$ (Fe,Co)N, which is paramagnetic. FeCoN nanodots treated with – 25 V show more amorphous structure with embedded nanocrystalline areas, especially visible at the bottom part of the film. The latter are characterized by the absence of the aforementioned (Fe,Co)N non-magnetic phase and by appearance of two new interplanar distances: 1.10 Å (green circles, which is consistent with planes from $Fm\bar{3}m$ Co, $Im\bar{3}m$ Fe or $Pm\bar{3}m$ FeCo), and 2.15 Å (pink circles, that matches planes from $Fm\bar{3}m$ Co, $Im\bar{3}m$ Fe, $Pm\bar{3}m$ FeCo or $Pm\bar{3}m$ $(Fe,Co)_4N$, all of them ferromagnetic). These findings agree with the aforementioned description of how ferromagnetic behavior emerges in the voltage-treated dots.

**Micromagnetic modelling of magneto-ionic vortices**

The observed evolution of magnetic properties upon gating was modelled using micromagnetic simulations. **Fig. 5** shows simulations of the magnetization reversal processes while applying – 25 V. As described in the Methods section, the thicknesses were chosen based on experimentally estimated values, and the magnitudes of $M_S$ and $K_u$ were tuned to obtain coercivities and saturation fields similar to experimental ones (**Fig. 2**). Importantly, an increase of both $M_S$ and $K_u$ with increasing thickness is needed to match the experimental loop characteristics (see **Supplementary Figs. 4,5** for details). This is consistent with a progressive loss of N content in the nanodots (in our case caused by voltage), as reported in the literature for FeCoN[37]. For the thinnest disks (**Fig. 5a**) the reversal process occurs *via* quasi-coherent rotation (with minor curling prior to saturation) and the loop shape and $H_C$ of 50 Oe agree well with the experimental hysteresis loop (**Supplementary Fig. 2b**). For intermediate and large thicknesses (**Fig. 5b,c**), vortex reversal is observed, also in agreement with the experimental results. Note that the concomitant increase of $H_C$ and $H_A$ with gating time is captured by the simulations. The high-coercivity square hysteresis loop measured after long-time gating at + 25 V (**Supplementary Fig. 2g**) was also modelled. For this, a high $K_u$ was used compared to the value for 2 nm thick disks initially formed at – 25 V. During the + 25 V treatment, the nitrogen front will move from the top to the bottom of the disks, resulting in a thinner ferromagnetic layer, but with material parameters similar to those of the previously generated thickest dots. Besides the high anisotropy, we slightly decreased the inter-grain exchange coupling to 30% to match the experimental value of $H_C$ (**Supplementary Fig. 6**). This is plausible since nitrogen ions might preferentially diffuse along grain boundaries and, in addition, grains might become less interconnected at progressively smaller thicknesses.

**Outlook**

We have shown that voltage can serve as an energy-efficient actuation method for tuning magnetization reversal mechanism and magnetic properties of FeCoN nanodots between paramagnetic, SD and magneto-ionic vortex states. Under the action of electric field, $N^{3-}$ ions diffuse *via* planar front formation, enabling the thickness of formed magnetic counterpart to gradually increase. This migration front, not present in other magneto-ionic systems, not only allows control over the magnetization reversal mechanism but also provides the capability to adjust magnetization and critical fields on-demand. The analog tuning of the vortex state, as demonstrated here, offers unique advantages over alternative methods like strain-mediated actuation relying on ferroelectric substrates, spin-transfer torque effects, or coupling with adjacent antiferromagnets, where the magnetization amplitude is held constant. In this context, vortions hold the potential to introduce innovative concepts in neuromorphic computing. Specifically, if integrated within spin-transfer-torque nano-oscillators—employed to emulate neurons[41]—the magneto-ionic vortex is poised to offer tunable magnetic synaptic weights through voltage-driven modulation of magnetization. Furthermore, alteration of magnetic anisotropy within the nanodots, as also demonstrated in this study, could impact the gyroscopic frequency of precessing vortex cores.[42] This, in turn, could influence the magnetization dynamics of neighboring oscillators, akin to the way in which spikes affect post-synaptic neurons, therefore further expanding the applicability of magneto-ionic vortices in neuromorphic spintronics systems and multi-state analog computing.

**Methods**

**Sample preparation**

FeCoN nanodots with a diameter of 280 nm and thickness of 35 nm were prepared by electron beam lithography (EBL) and subsequent magnetron sputtering. EBL was performed by using a Raith GmbH-150-TWO nanofabrication system. A double layer MMA/PMMA photoresist mask was deposited by spinning for 55 s at 1500 rpm and baked for 60 s at 180 ºC on top of [100]-oriented Si substrates previously coated with Ti (20 nm)/Pt (50 nm) layers (deposited by sputtering). Photoresists were developed using 1:3 methyl isobutyl ketone/isopropanol (MIBK/IPA) for 60 s. After pattering, 35 nm of ternary nitride $Fe_{0.65}Co_{0.35}N$ (FeCoN) was deposited by reactive magnetron co-sputtering at room temperature. An AJA International ATC 2400 sputtering system with a base pressure of around $8\times10^{-8}$ Torr was used. The target-to-substrate distance was kept at 11 cm. Reactive sputtering was done at a total pressure of $3\times10^{-3}$ Torr in an $Ar/N_2$ mixed atmosphere. During the sputtering, the $Ar:N_2$ flow ratio was set to 1:1 in order to produce nitrogen-rich nanodots exhibiting paramagnetic behaviour. Based on a previous study, the metallic Fe target was operated at a constant DC power of 50 W, while the Co target was channelled to a RF source with a power of 55 W[33]. Finally, non-exposed photoresist was removed by a lift-off process in acetone using an ultrasonic bath, revealing FeCoN nanodots.

**MOKE measurements**

*In-situ* measurements of hysteresis loops during voltage actuation were acquired using a Magneto-Optical Kerr Effect (MOKE) magnetometer (NanoMOKE3, Durham Magneto Optics

Ltd.) in a longitudinal geometry with the magnetic field applied in-plane. The system enables the application of time-varying magnetic fields to the sample, and the rotation of the polarization plane of the reflected laser beam, referred to as the Kerr signal, is measured as a function of the applied field and plotted in the form of the hysteresis loops. For *in-situ* measurements in a propylene carbonate (PC) liquid electrolyte, we affixed a custom-made quartz electrochemical cell onto the sample holder. The cell, which we designed specifically for the magneto-ionic measurements in liquid, housed the Si/Ti(20 nm)/Pt(50 nm)/FeCoN nanodots sample and featured electrical contacts to the bottom Pt layer (working electrode) and Pt wire (counter electrode) in a capacitor configuration. We applied a gate voltage ($V_G$) between the mentioned electrodes using an external Agilent B2902A power supply, while simultaneously measuring the magnetic response of the sample, induced by the migration of ions within the FeCoN nanodots.

**Magnetic force microscopy (MFM) measurements**

Magnetic Force Microscopy (MFM) measurements were conducted on both as-grown and voltage-treated samples using the MFP-3D Origin+ Atomic Force Microscope from Asylum Research, Oxford Instruments. The measurements were carried out in two-pass mode, employing previously magnetized ASYMFM-R2 probes. For the second pass, we used a lift-height of 15 nm. Variable magnetic fields were applied in an in-plane geometry using the Variable Field Module provided by the same manufacturer.

**Electron microscopy**

Scanning electron microscopy (SEM) imaging was performed using a Zeiss Merlin SEM microscope. High-resolution transmission electron microscopy (HR-TEM), high-angle annular dark-field scanning transmission electron microscopy (HAADF-STEM), and electron energy loss spectroscopy (EELS) were performed on a FEI TECNAI G2 F20 HRTEM/STEM microscope with a field emission gun operated at 200 kV. Cross-sectional lamellae of the samples were cut by focused ion beam after the deposition of Pt protective layers and were subsequently placed onto a Cu TEM grid.

**Micromagnetic simulations**

Micromagnetic simulations of the magnetization reversal process were conducted for disks with a diameter of 280 nm using MuMax3[43,44]. Cells with $d_x = d_y = 2.1875$ nm and $d_z = 2$ nm (along the thickness direction) were used to simulate nanodots with thicknesses ranging from 2 to 12 nm. Grains were included[45], approximately 5 nm in size, since the disks are polycrystalline, and the grains are needed to reproduce the wide experimental hysteresis loop shapes (see Supplementary Information section II for details). A random anisotropy direction was assigned to each grain, and the anisotropy value assigned to each grain was allowed to vary randomly in magnitude by 10% from the chosen $K_u$ value. The hysteresis loops shown in Fig. 5 are averaged over 10 runs, where a new set of randomized anisotropy parameters was chosen for each run to capture the effect of averaging over an array. The exchange constant was held fixed at 1.3 μerg/cm, and the exchange coupling between grains was reduced by 10%. To reduce the run time, only the downward leg of the hysteresis loop was simulated. In all cases a constant $M_S$ is used across the thickness, and it should be considered as an effective $M_S$ that might stem from the phases with different N content. The thicknesses for the hysteresis loops shown in Fig. 5 were chosen to match experimentally estimated values of the thickness of the magnetic layer for selected voltage treatments, and then the $M_S$ and $K_u$ parameters were tuned to obtain hysteresis loop shapes that matched the experiment. The parameters used for the simulations shown in Fig. 5 are: (a) $L = 2$ nm, $M_S = 400$ emu/cm$^3$, $K_u = 2.5 \times 10^5$ erg/cm$^3$ to represent the thinnest disks after a negative voltage treatment; (b) $L = 8$ nm, $M_S = 500$ emu/cm$^3$, $K_u = 3.5 \times 10^5$ erg/cm$^3$ for an intermediate voltage treatment; and (c) $L = 12$ nm, $M_S = 550$ emu/cm$^3$, $K_u = 6 \times 10^5$ erg/cm$^3$ for the thickest disks. Note that the $M_S$ and $K_u$ values increase while increasing the thickness of the voltage-induced ferromagnetic disks, which is consistent with a progressive denitriding process of FeCoN[37,38]. Magnetic force microscopy images were also calculated, where the MFM contrast was obtained by finding the gradient of the out-of-plane component of the stray magnetic field at a height of 50 nm above the magnetic disk. The calculated images were convolved with a Gaussian function with a width parameter of 50 nm to account for the finite width of the MFM tip. Additional simulations, including simulations conducted to determine the parameters needed to reproduce the experimental hysteresis loops after long-term positive voltage treatment, are included as Supplementary Materials. The parameters $L = 2$ nm, $M_S = 500$ emu/cm$^3$ (slightly lower than the value for the thickest disks) and $K_u = 6 \times 10^5$ erg/cm$^3$ (the same value used for the thickest disks) were used to represent the thinnest disks after a positive long-term voltage treatment. The exchange coupling between grains was set to 30% of the underlying exchange value inside the grains.

## Data availability

All data are available in the main text or Supplementary Information. Source data are provided with this paper.

## Acknowledgements

This work has been supported by the European Research Council (2021-ERC-Advanced REMINDS Grant Nº 101054687). Partial financial support from the Generalitat de Catalunya (2021-SGR-00651), the Spanish Government (PID2020-116844RB-C21) and (PID2021-124680OB-I00) funded by MCIN/AEI/10.13039/501100011033 by ERDF "A way of making Europe" is also acknowledged. A.B. acknowledges support from MICIN Predoctoral Fellowship (PRE2019-09781). K.S.B. acknowledges support from the W.M. Keck Foundation. We thank Dr. Cristina Navarro-Senent for her assistance with graphical design.

## Author information

### Contributions

J.S. conceived and supervised the study. I.S., A.P. and Z.M. prepared the samples, with the help of A.B. I.S. and J.S. performed *in situ* MOKE measurements during sample gating and analyzed the results. I.S. and J.S. performed HR-TEM, HAADF-STEM and EELS measurements and analyzed the results. I.S. performed and analyzed the MFM measurements. K.S.B. conducted the micromagnetic simulations with the inputs from J.S. and I.S. I.S. wrote the manuscript with inputs from J.S. and K.S.B. All authors contributed to discussions and editing.

### Corresponding authors

Correspondence to Irena Spasojevic and Jordi Sort.

### Ethics declarations

The authors declare no competing interests.

# Figures

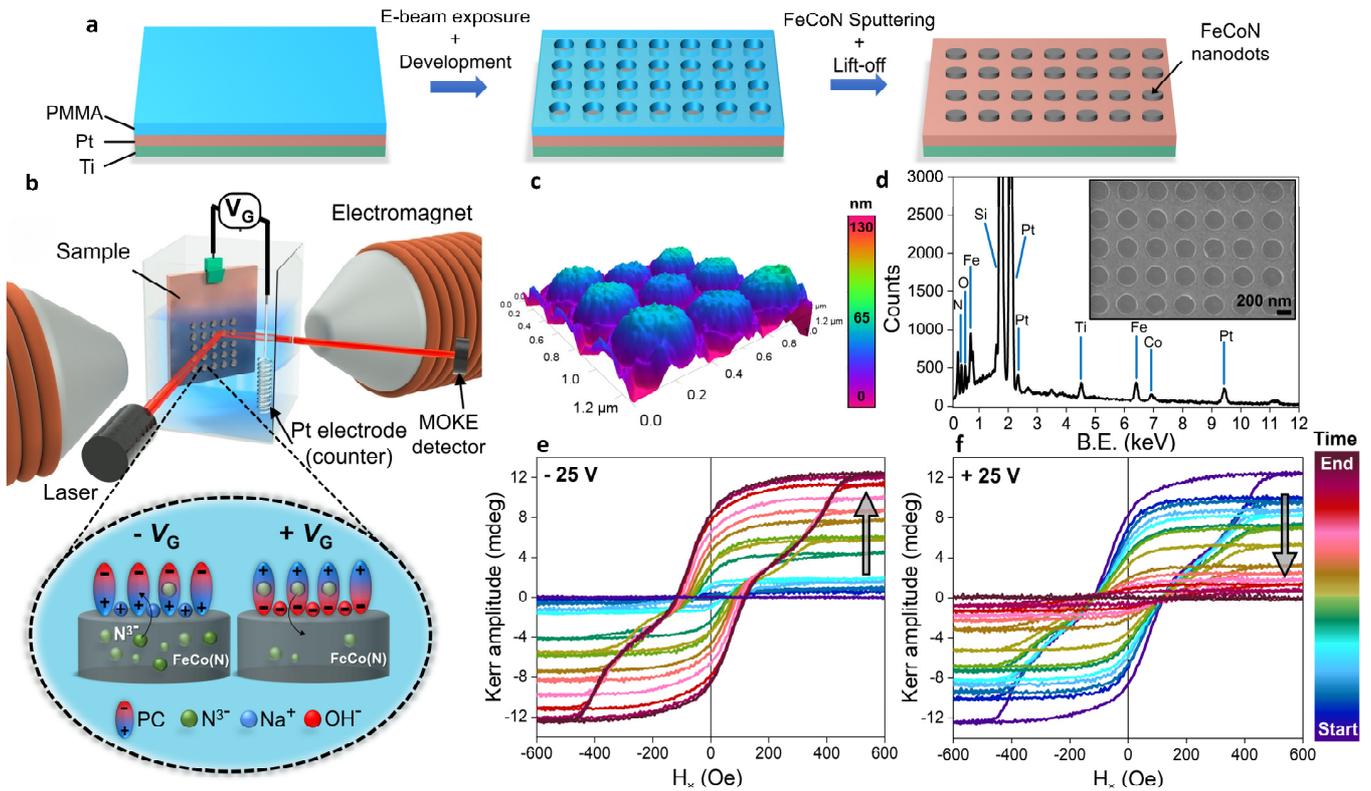

**Fig. 1. Device and set-up schematics for voltage-induced ferromagnetism in FeCoN nanodots. a**, Schematic representation of FeCoN nanodots preparation by electron-beam lithography and sputtering. **b**, Schematics of the electrochemical cell, custom-made in our lab, used for the *in-situ* measurements of voltage-induced magnetism within FeCoN nanodots by MOKE (see Methods). A zoomed-in panel depicts electric-double layer formation at the FeCoN/PC electrolyte interface and $N^{3-}$ migration out of and into the FeCoN layer when negative or positive gate voltage is applied between working and counter electrode, respectively. **c**, 3D topography of FeCoN nanodots measured by AFM in tapping mode. **d**, EDX analysis of as-grown FeCoN nanodots. The inset shows an SEM image of the nanodots array. **e, f,** Evolution of hysteresis loops measured by MOKE during negative (**e**) and positive (**f**) voltage actuation of FeCoN nanodots. Starting from the paramagnetic state (purple curve in **e**), Kerr amplitude progressively increases over time. Simultaneously, the measured hysteresis loops undergo a transition in shape, evolving from square-like to constricted (see Supplementary Fig. 2). The opposite happens when positive voltage is applied, whereby the initial paramagnetic state can be reversibly restored.

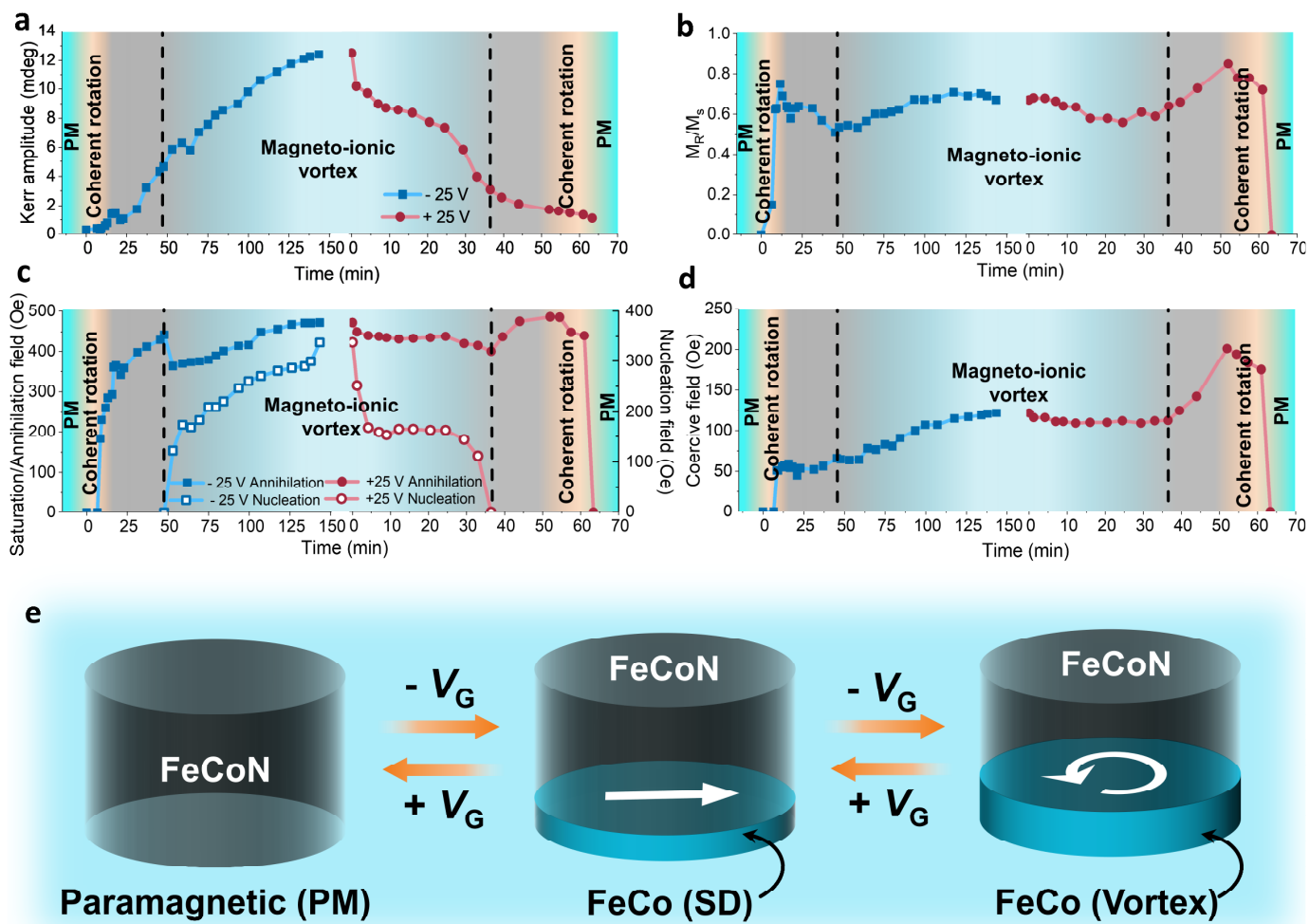

**Fig. 2. Magneto-ionic vortex (*i.e.*, vortion) operation.** Evolution of hysteresis loop parameters and main reversal mechanisms with voltage actuation time such as Kerr amplitude measured at saturation ($H_{applied}$ = 2000 Oe) (**a**), squareness ratio $M_R/M_S$ (**b**), saturation/annihilation field (left axis) and nucleation field (right axis) (**c**) and coercivity (**d**). Blue lines (denoted by full or empty squares) and red lines (denoted by full or empty circles) depict evolution of a given parameter with time when negative or positive voltage is applied, respectively. For short actuation time, a thin ferromagnetic layer (highly or fully depleted in nitrogen) is formed, resulting in a low Kerr amplitude and square-like hysteresis loops. This indicates presence of SD state and magnetization reversal by coherent rotation (orange regions). Conversely, for extended actuation at – 25 V, the ferromagnetic layer thickness increases gradually, which stabilizes the magneto-ionic vortex state and provides a means to fine-tune its magnetization amplitude and critical fields (blue regions). The grey regions indicate transient states, representing the transitional phases between the SD and vortex states, and *vice versa*. The black dotted vertical lines delimitate the regions where the magneto-ionic vortex state is fully stable. **e** Schematic representation of the transition between paramagnetic (PM), SD and magneto-ionic vortex states within the same FeCo(N) nanodot driven by $N^{3-}$ ion migration.

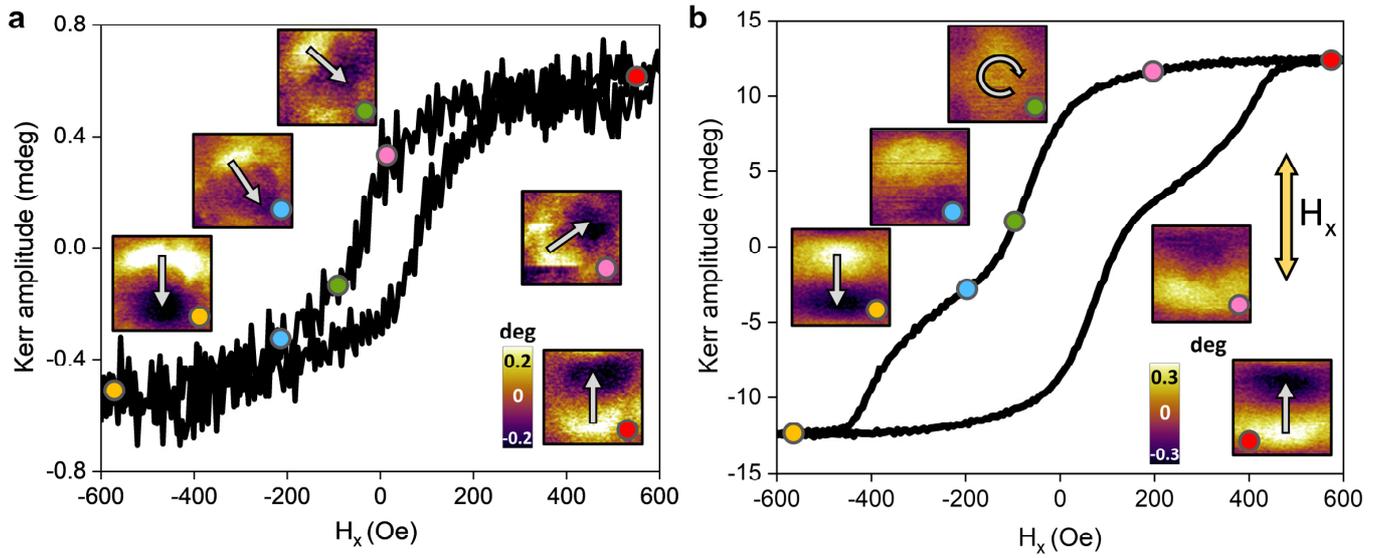

**Fig. 3. MOKE hysteresis loops with MFM imaging for selected voltage-actuation times. a,** Hysteresis loop and correspoding MFM phase images of one of the disks measured after short-time voltage actuation (– 25 V), corresponding to the formation of thin ferromagnetic layer. The shape of the hysteresis loop, together with the observed rotation of dipolar contrasts, shows that in such thin magnetic layers, magnetization reversal proceeds *via* coherent rotation. **b,** Hysteresis loop and correspoding MFM phase images measured after long-time voltage actuation, corresponding to the formation of a thick ferromagnetic layer. In this case, the obtained results reveal that magnetization reversal occurs through formation of a swirling spin (*i.e.*, vortion) state.

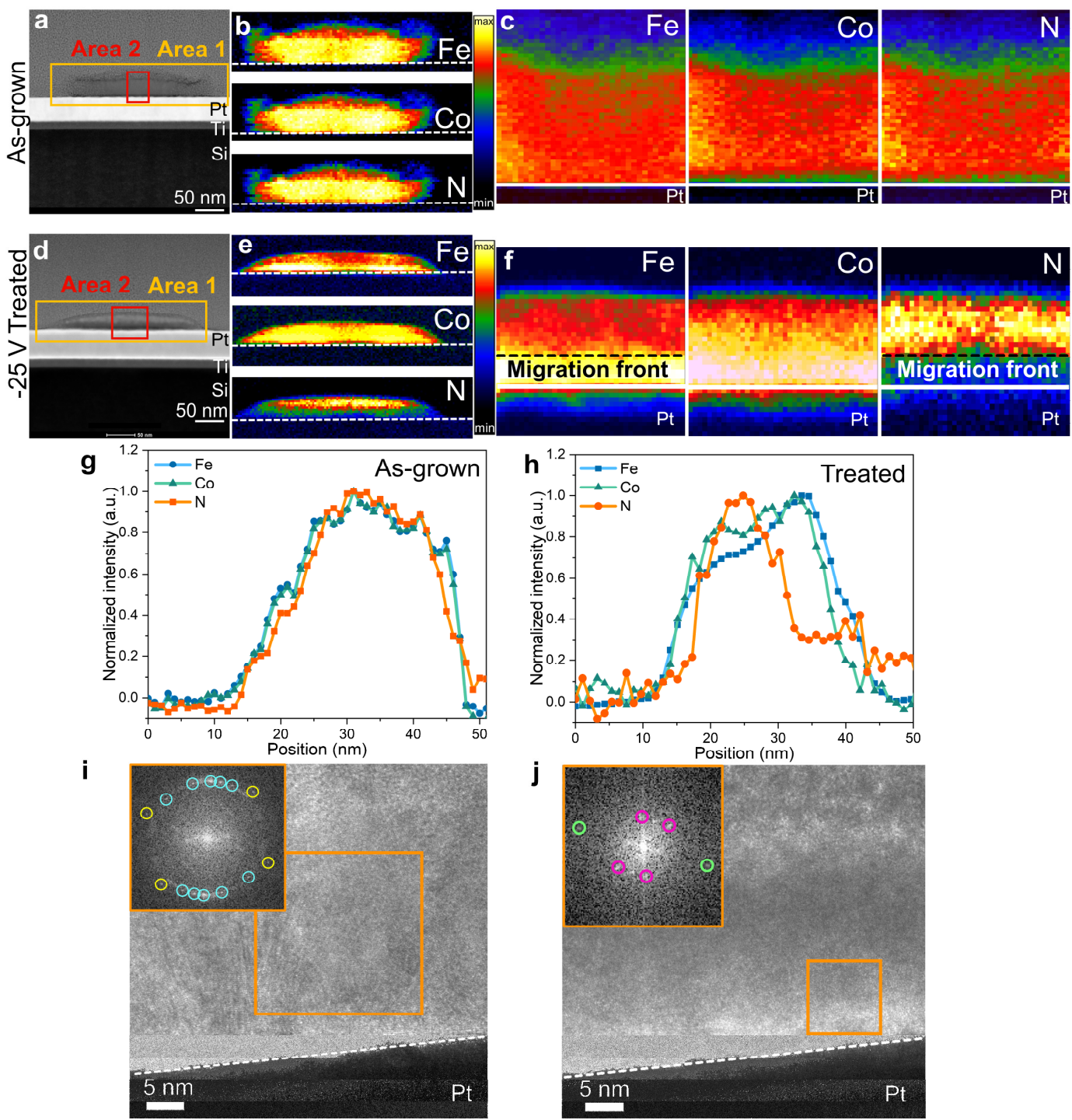

**Fig. 4. HR-TEM and EELS compositional mappings. a**, HAADF-STEM image of as-grown FeCoN nanodot. Red and orange rectangles indicate two areas where EELS compositional mapping was performed. EELS compositional maps, comprising the entire as-grown nanodot (Area 1, **b**) and closer examination of nanodot's interior (Area 2, **c**), show uniform distribution of Fe, Co and N throughout both regions. **d**, HAADF-STEM image of FeCoN nanodot treated with − 25 V. EELS compositional maps performed in two areas indicated in panel **d** show that upon negative voltage treatment Fe/Co ions and $N^{3-}$ ions migrate in opposite directions, thereby creating a distinct planar migration front indicated by black dashed lines. This planar front divides the nanodots into two sub-layers with distinct magnetic properties: ferromagnetic

FeCo/(Fe,Co)$_4$N (*i.e.*, nitrogen-depleted) at the bottom part and paramagnetic FeCoN (*i.e.*, nitrogen-rich) in the upper part of the nanodot. **g, h**, Normalized EELS compositional line profiles of as-grown and treated FeCoN dots, respectively, collected starting from the top of the nanodot. **i, j**, HR-TEM images of nanodot's interior before (**i**) and after (**j**) voltage treatment (at – 25 V), including the FFT of the areas marked by orange squares. Interplanar distances marked by yellow and cyan circles in **i** correspond to the interplanar distances of 2.26 Å, 2.60 Å, respectively, and stem from (200) and (111) planes of $F\bar{4}3m$ (Fe,Co)N. After voltage treatment (**j**), new interplanar distances of 1.10 Å (green) and 2.15 Å (pink) emerge. Interplanar distance 1.10 Å can be ascribed to (311) planes of fcc $Fm\bar{3}m$ Co, (211) planes of body-centered cubic (bcc) $Im\bar{3}m$ Fe, or to (211) planes of simple cubic $Pm\bar{3}m$ FeCo alloy. Similarly, interplanar distance 2.15 Å can be related to (111) planes of fcc $Fm\bar{3}m$ Co, (110) planes of bcc $Im\bar{3}m$ Fe, (110) planes of $Pm\bar{3}m$ FeCo or (111) planes of $Pm\bar{3}m$ (Fe,Co)$_4$N.

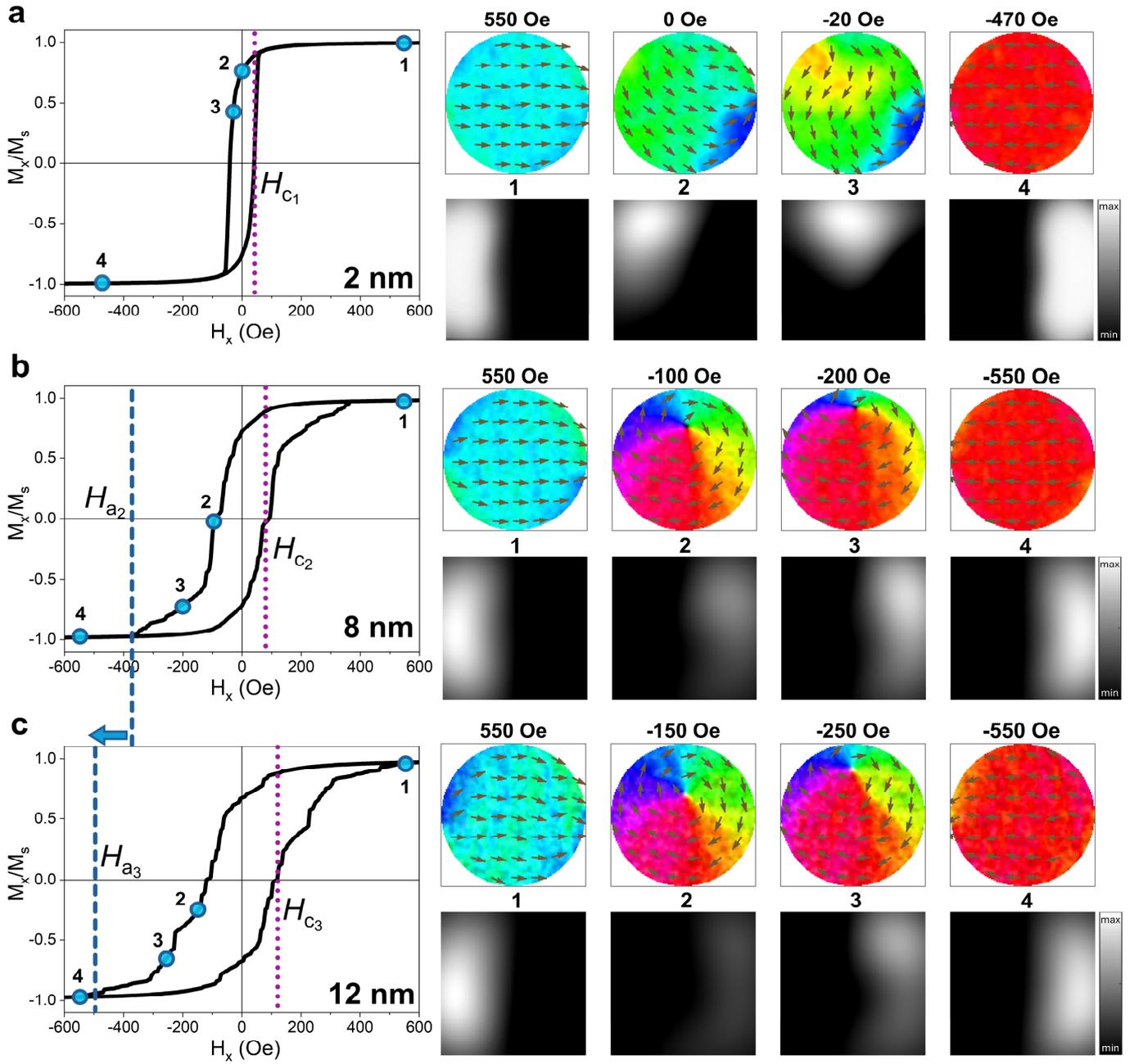

**Fig. 5. Micromagnetic simulations showing the reversal process of the magnetic disks upon application of – 25 V**. Simulated hysteresis loops are shown for 280-nm diameter magnetic disks where the thicknesses and magnetic parameters were chosen to capture the reversal behavior of (a) a short-term voltage treatment (thickness 2 nm, $M_S$ = 400 emu/cm$^3$, $K_u$ = 2.5× 10$^5$ erg/cm$^3$), (b) a mid-term voltage treatment (thickness 8 nm, $M_S$ = 500 emu/cm$^3$, $K_u$ = 3.5 × 10$^5$ erg/cm$^3$), and (c) long-term voltage treatment (thickness 12 nm, $M_S$ = 550 emu/cm$^3$, $K_u$ = 6 × 10$^5$ erg/cm$^3$). The hysteresis loops shown on the left are an average of 10 runs (see Supplementary Fig. 5 for details), and the images on the right show spin distributions (top) and calculated MFM images (bottom) at selected fields for a representative run. Blue and purple dashed lines mark the values of annhilation and coercive fields, respectively.

# Supplementary Information

## Magneto-Ionic Vortices: Voltage-Reconfigurable Swirling-Spin Analog-Memory Nanomagnets

Irena Spasojevic[1,*], Zheng Ma[1], Aleix Barrera[2], Federica Celegato[3], Ana Palau[2], Paola Tiberto[3], Kristen S. Buchanan[4], Jordi Sort[1,5,*]

[1]Departament de Física, Universitat Autònoma de Barcelona, 08193 Bellaterra, Spain
[2]Institut de Ciència de Materials de Barcelona (ICMAB-CSIC), Campus UAB, Bellaterra 08193, Barcelona, Spain
[3]Advanced materials and Life science Divisions, Istituto Nazionale di Ricerca Metrologica (INRIM), Strada delle Cacce 91, 10135 Turin, Italy
[4]Department of Physics, Colorado State University, Fort Collins, Colorado 80523, USA
[5]ICREA, Pg. Lluís Companys 23, 08010 Barcelona, Spain

[*]Email: Irena.Spasojevic@uab.cat, Jordi.Sort@uab.cat

## I. Additional experimental results

Figure S1 shows the transmission electron microscopy image of a cross section of the array of nanodots in the as-grown state, to complement the observations provided in Fig. 4 of the manuscript. The nanodots have a diameter of 280 nm and thickness of around 35 nm.

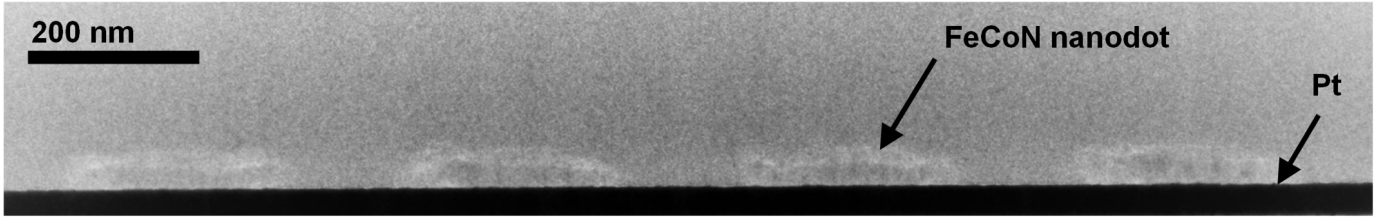

**Figure S1.** Transmission electron microscopy (TEM) image of cross-sectional lamella containing as-grown FeCoN nanodots.

Figure S2 shows the detailed evolution of the shape of the hysteresis loops upon gating using a negative voltage (panels a – d) and, subsequently, positive voltage (panels e – h). Starting from a paramagnetic state (a), denitriding of (Fe,Co)N (for $V_G = -25$ V) results in a progressive increase of the Kerr signal amplitude because of the formation of ferromagnetic phases such as FeCo or (Fe,Co)$_4$N. Remarkably, for short actuation times, the shape of the hysteresis loops is rather square (panel b), suggesting reversal of a single-domain (SD) state via coherent rotation. Longer actuation times result in constricted hysteresis loops, typical of magnetization reversal *via* vortex-state formation (panel d). For $V_G = +25$ V, the shape of the loops evolves from constricted (panels e,f, vortex state) to square-like (panel g, SD state) and then to paramagnetic (panel h).

While the evolution of the hysteresis loops' shapes is symmetric for negative and positive voltages, one can also observe that the coercivity progressively increases when going from SD to vortex state and *vice versa*, giving certain asymmetry to the values of $H_C$. Namely, the final coercivity values during positive gating ($\approx 170$ Oe) are substantially higher compared to initial values ($\approx 50$ Oe) when negative voltage is applied. The asymmetry in coercivity behavior may be caused by the progressive change in the magnetic anisotropy during the gating. Namely, starting from paramagnetic (Fe,Co)N phase, upon negative gating, nitrogen-depleted, low-anisotropy (and low $H_C$) ferromagnetic phases are formed first at the bottom of the nanodots, giving overall low coercivity values. With prolonged gating time, nitrogen-free, higher anisotropy phases such as FeCo also form at the bottom of the nanodots. Note that the magnetic anisotropy of FeCoN alloys increases with a decrease of the N content [1,2]. The formation of these phases subdivides the nanodots into two sub-layers with clearly dissimilar nitrogen content. These layers are delimited by the planar ion migration front, with the bottom sub-layer being ferromagnetic and exhibiting variable thickness depending on the actuation time. This is evidenced in Fig. 4 of the manuscript. In the reverse scenario (positive gating), voltage causes the reintroduction of $N^{3-}$ ions from the electrolyte to the nanodots. Given that the Kerr amplitude decreases over time, it is plausible to assume that the planar migration front will progressively move downwards. The coercivity remains high because nitrogen ions do not immediately reach the bottom of the nanodots. Consequently, the bottom sub-layer, with progressively lower thickness, retains the high-coercivity ferromagnetic phases (very much depleted in N), which have higher anisotropy.

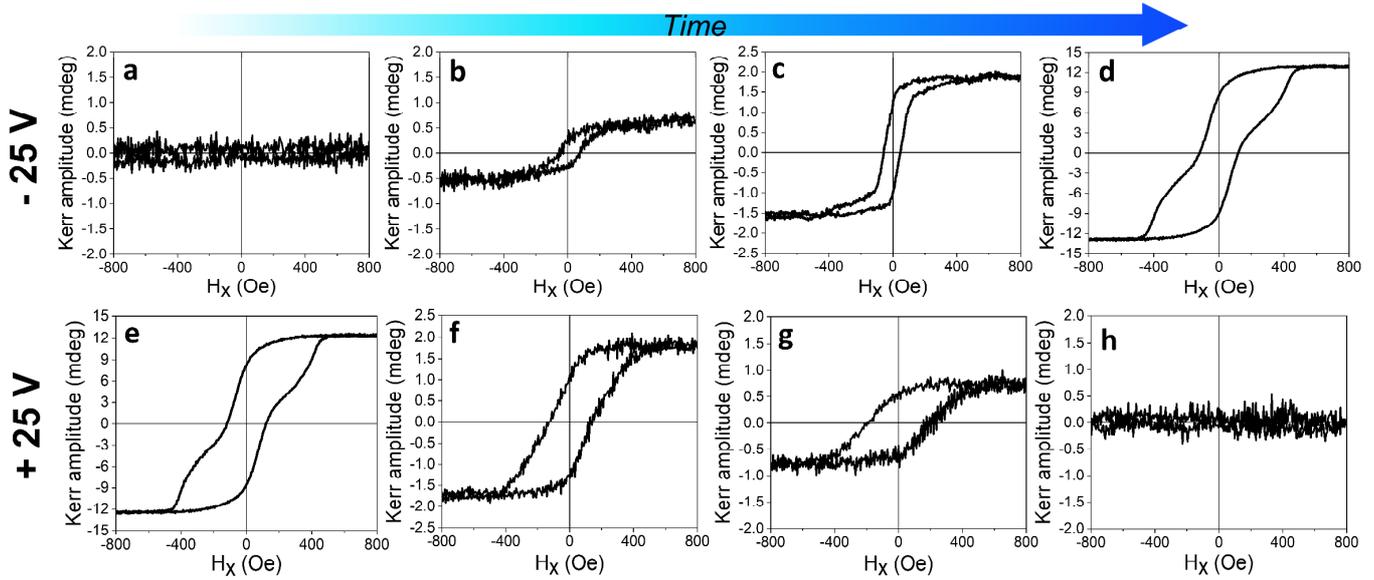

**Figure S2.** *In-situ* **measured MOKE hysteresis loops during voltage-actuation of FeCoN nanodots.** Measured MOKE hysteresis loops, starting from pure paramagnetic behavior (**a**) during the application of – 25 V, provide evidence of distinct spin configurations and reversal mechanisms over time. These include the single domain state and magnetization reversal via coherent rotation (**b**), a transient-like state where both single domain and vortex states may coexist (**c**), and finally, the fully stable vortex state (**d,e**). Conversely, application of + 25 V leads to the reversible transition to transient-like (**f**), single domain (**g**) and paramagnetic (**h**) states.

Figure S3 provides magnetic force microscopy (MFM) images that complement those shown in Fig. 3 of the manuscript. In the presence of external magnetic field *H* = 700 Oe (which is high enough to ensure magnetic saturation), no magnetic contrast is observed in the as-grown nanodots, whereas clear dipolar contrast emerges in all dots after treatment with $V_G$ = – 25 V.

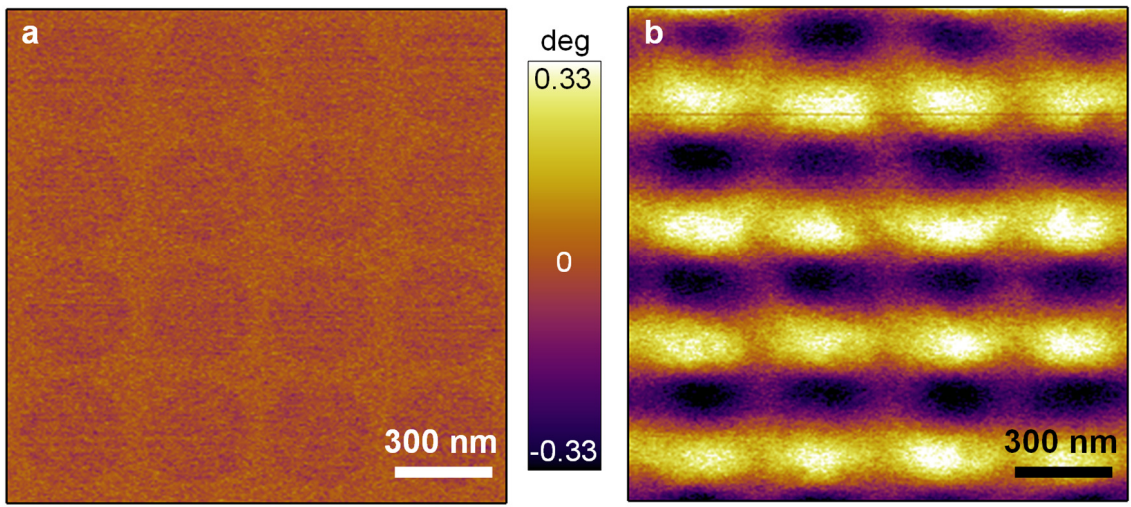

**Figure S3. Magnetic Force Microscopy (MFM) characterization of as-grown and voltage-treated FeCoN nanodots**. MFM phase images of the as-grown FeCoN nanodots (**a**) and those treated with -25 V (**b**), imaged in the presence of external magnetic field $H$ = 700 Oe. In the as-grown FeCoN nanodots, faint contrast is observed, primarily stemming from topographic crosstalk, resulting in the absence of a discernible magnetic signal. This observation aligns with the findings measured by MOKE (see **Fig. S2a**).

## II. Additional details on the micromagnetic simulations

Additional micromagnetic simulations were performed to understand the roles of the sample thickness, the inclusion of grains, and the magnitude of magnetic anisotropy inside the grains. Micromagnetic simulations of the magnetization reversal process were conducted for disks with a diameter to 280 nm using MuMax3 [3,4]. Cells with $dx = dy = 2.1875$ nm and $dz = 2$ nm (in the thickness direction) were used to simulate disks with thicknesses $L$ = 2 - 12 nm. In all cases the exchange constant was held fixed at 1.3 μerg/cm (a typical value for Fe-Co nitrides [5,6]). The downward leg of the hysteresis loop was simulated by first relaxing the magnetic disk at the maximum field, and spin distributions at subsequent field steps were obtained through energy minimization at each field step.

Figure S4 shows single-run hysteresis loops obtained for the 280-nm diameter disks with a constant $M_S$ value of 550 emu/cm$^3$. As shown in Fig. S4a, when anisotropy is neglected and the magnetic properties of the disk are uniform (*i.e.*, no grains are included), the hysteresis loops show the classic double lobe shape that is associated with a vortex reversal process for $L > 4$ nm. For the thinnest disks, $L \leq 4$ nm, the reversal process is quasi-SD (coherent rotation). The annihilation field for the vortex increases with increasing $L$, which is expected [7]. The $M_S$ value used in the simulations was chosen to obtain an annihilation field similar to the experimental value for the largest $L = 12$ nm. The experiments indicate that (Fe,Co)N (paramagnetic) progressively loses nitrogen and is first transformed to ferromagnetic phases with low anisotropy and low $M_S$, and then to FeCo (with higher effective anisotropy and $M_S$). In the simulations, the $M_S$ value is low compared to that of bulk FeCo alloys, however it should be viewed as an effective, thickness-averaged $M_S$ across a layer that reflects a range of $M_S$ values from phases with different N content. The hysteresis loops shown in Fig. S4a are consistent with the experimental observation of vortex reversal processes for the thickest disks and quasi-SD reversal for the thinnest, however, the vortex loops in Fig. S4a show zero coercivity, whereas the loops measured experimentally show coercivities $H_C$ of 50-170 Oe.

Next, grains were included to see if this produces loop shapes that are closer to the experimental hysteresis loops. Figures S4b,c show the effect of including grains [8] that are 5 nm in size. A uniaxial anisotropy value $K_{u,i}$ and an anisotropy direction is assigned to each of the grains where the anisotropy magnitude $K_{u,i}$ is varied by 10% about a mean value $K_u$, and the anisotropy direction is randomized. The exchange coupling between grains was set to 90% of the underlying exchange value. In Fig. S4b, the same $K_u = 4 \times 10^5$ erg/cm$^3$ is used for all thicknesses, and in Fig. S4c, the $K_u$ values decrease with decreasing $L$ ($K_u = (6.0, 5.2, 4.5, 4.0, 3.4, 3.0) \times 10^5$ erg/cm$^3$. Note that this range of $K_u$ values is plausible for FeCoN compounds [9]. The reversal process still proceeds by vortex nucleation and annihilation for larger $L$ and by quasi-SD rotation for smaller $L$, however, the coercivity is now non-zero for all $L$, which is in better agreement with the experimental results. In the experiments, $H_C$ increases with increasing $L$ for the negative voltage treatment and, based on the results shown in Fig. S4b,c, this suggests that an increase in $K_u$ with increasing $L$ is required to reproduce the trend in $H_C$ observed in the experiments.

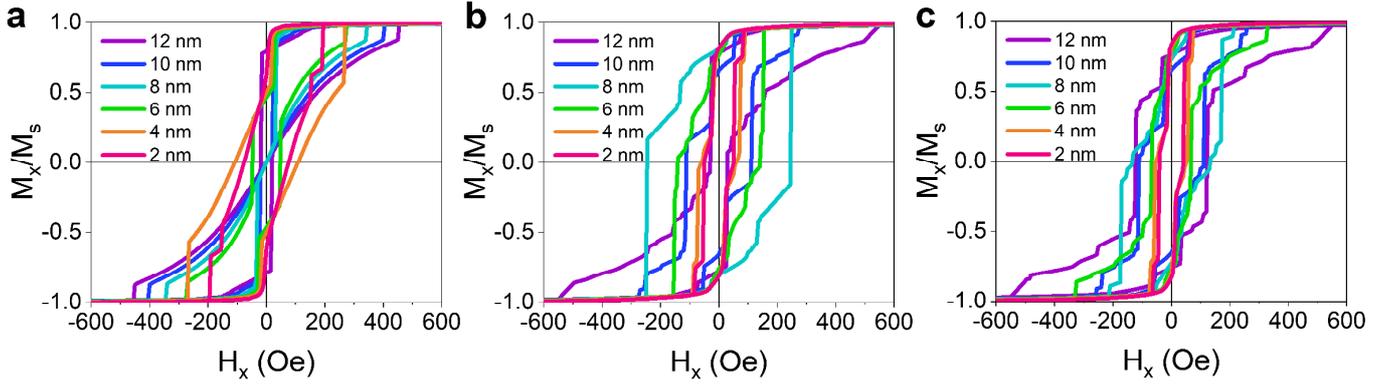

**Figure S4.** Simulated single-run hysteresis loops obtained for 280-nm diameter disks with $M_S$ = 550 emu/cm$^3$ for $L$ of 12 to 2 nm (indicated in the legend) with (a) no anisotropy and no grains, (b) 5-nm grains with $K_u = 4 \times 10^5$ erg/cm$^3$, and (c) 5-nm grains with decreasing $K_u$ values following a decrease in $L$ ($K_u = (6.0, 5.2, 4.5, 4.0, 3.4, 3.0) \times 10^5$ erg/cm$^3$), where the highest $K_u$ corresponds to $L = 12$ nm).

In Fig. S4, a constant $M_S$ value was used in order to isolate the effects of adding grains and the role of the magnitude of $K_u$. However, experimental results from the literature suggest that a lowering of the $K_u$ magnitude should be accompanied with a reduction in $M_S$ (since both parameters decrease in FeCoN with an increase of the N content [1,2]). Furthermore, the single-run hysteresis loops in Fig. S4b,c show multiple abrupt steps, whereas the experimental loops are much smoother. To obtain a closer match between the experimental results and the simulations, the simulations were repeated with 10 runs and the average was compared with the experimental loop shapes. For each run, a new set of random values was used for the grain anisotropy and the hysteresis loops are averaged to approximate the response of an array of disks (note that the MOKE laser spot size is around 3 – 4 µm, so indeed, several nanodots are measured simultaneously). The $M_S$ and $K_u$ values were then tuned for selected thicknesses, $L = 2, 8$, and 12 nm, to determine the values that produce hysteresis loop shapes that most closely resemble the experimental results. The average hysteresis loops and the associated spin distributions upon gating using − 25 V are shown in the main manuscript, and Fig. S5a–c show the individual runs for each of the considered cases as well as the averaged loops.

The hysteresis loop for 2 nm thick magnetic disks was also simulated using the parameters obtained for the thickest (12 nm) disks after a negative voltage treatment to see if preserving these parameters can explain the larger $H_C$ observed after a voltage treatment with + 25 V. This approach leads to a value of $H_C$ of ~100 Oe (Fig. S5d), which is high compared to the coercivity in Fig. S5a but still low compared to the 170 Oe value observed experimentally. As shown in Fig. S6, maintaining the $K_u$ value from $L = 12$ nm and simultaneously reducing the inter-grain exchange coupling from 90% to 30% leads to a hysteresis loop with a larger $H_C$ and a loop shape that agrees well with the experiments after a voltage treatment of + 25 V. The $M_S$ value was also slightly reduced compared to $L = 12$ nm, though this has a minimal effect on the $H_C$ or the loop shape for $L = 2$ nm. Finally, the spin configurations and simulated MFM images corresponding to long-term voltage actuation with $V_G = + 25$ V are also provided in Fig. S6.

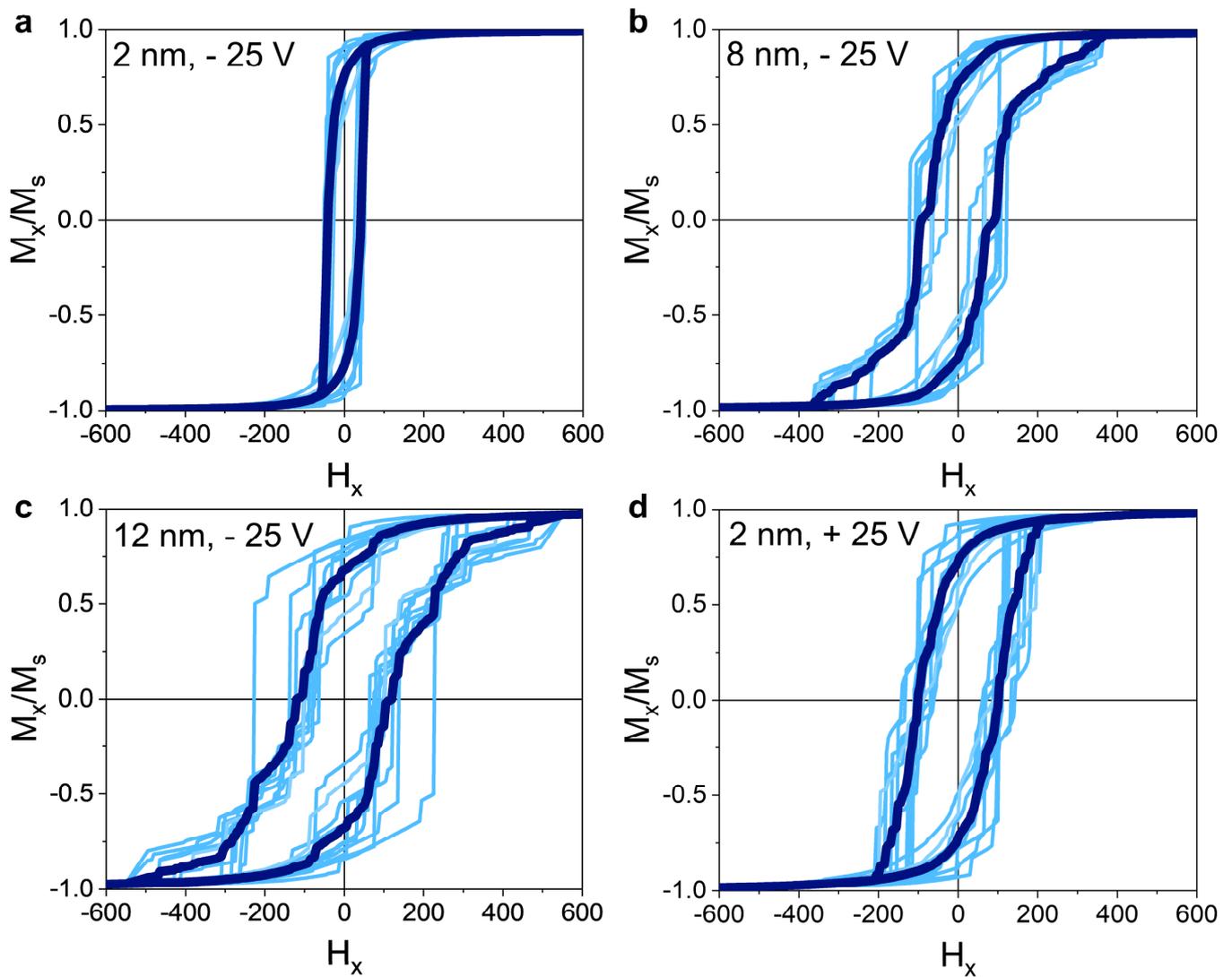

**Figure S5.** Hysteresis loops for 280-nm diameter disks showing the single-run hysteresis loops (10 runs) as well as the average over the 10 runs (dark blue line). The parameters were selected to reproduce experimental results for a (a) short negative voltage treatment ($L = 2$ nm, $M_S = 400$ emu/cm$^3$, $K_u = 2.5 \times 10^5$ erg/cm$^3$) (b) a mid-length negative voltage treatment ($L = 8$ nm, $M_S = 500$ emu/cm$^3$, $K_u = 3.5 \times 10^5$ erg/cm$^3$) (c) the longest negative voltage treatment ($L = 12$ nm, $M_S = 550$ emu/cm$^3$, $K_u = 6 \times 10^5$ erg/cm$^3$). Similar parameters to those used for $L = 12$ nm were then used to simulate the hysteresis loop expected for (d) a subsequent positive voltage treatment ($L = 2$ nm, $M_S = 500$ emu/cm$^3$, $K_u = 6 \times 10^5$ erg/cm$^3$). In all cases, the exchange coupling between grains was set to 90% of the underlying exchange value inside the grains (1.3 µerg/cm).

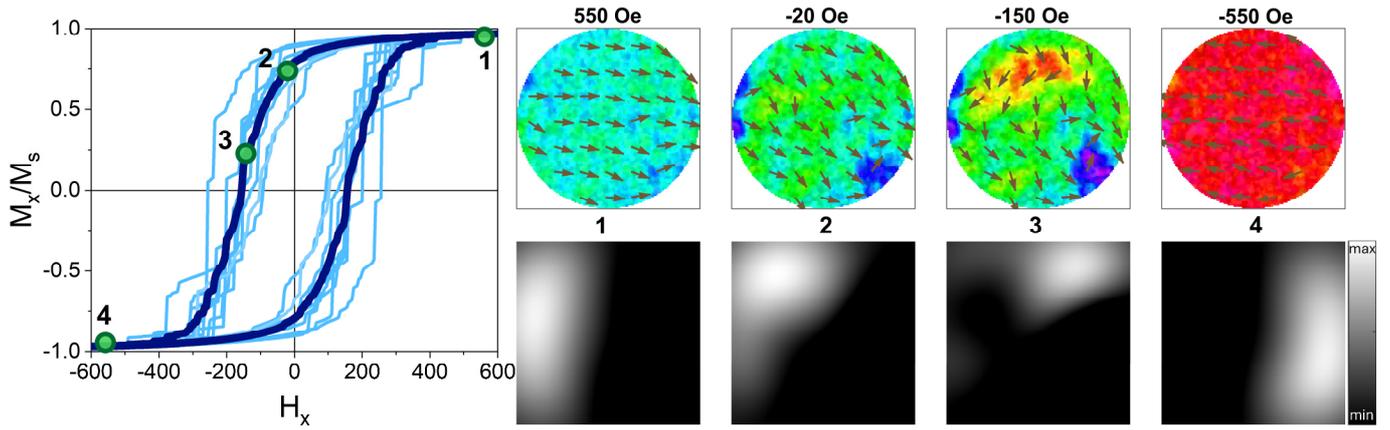

**Figure S6. Micromagnetic simulations showing the reversal process of the magnetic disks upon long-term application of + 25 V**. Simulated single-run hysteresis loops (10 runs), as well as averaged hysteresis loop (dark blue line) shown for 280-nm diameter magnetic disk with a thickness of 2 nm, $M_S$ = 550 emu/cm$^3$, $K_u$ = 6 × 10$^5$ erg/cm$^3$ and inter-grain exchange coupling of 30% of the underlying exchange value inside the grains. The images on the right show spin distributions (top) and calculated MFM images (bottom) at selected fields for a averaged run.